# Title: Search Still Matters: Information Retrieval in the Era of Generative AI

Corresponding Author:

William Hersh, MD
Professor
Department of Medical Informatics & Clinical Epidemiology
School of Medicine
Oregon Health & Science University
BICC
3181 SW Sam Jackson Park Rd.
Portland, OR, USA   97239
Voice: +1-503-494-4563
Email: hersh@ohsu.edu

Word count, excluding title page, abstract, references, figures and tables: 1876

**ABSTRACT**

Objective

Information retrieval (IR, also known as search) systems are ubiquitous in modern times. How does the emergence of generative artificial intelligence (AI), based on large language models (LLMs), fit into the IR process?

Process

This perspective explores the use of generative AI in the context of the motivations, considerations, and outcomes of the IR process with a focus on the academic use of such systems.

Conclusions

There are many information needs, from simple to complex, that motivate use of IR. Users of such systems, particularly academics, have concerns for authoritativeness, timeliness, and contextualization of search. While LLMs may provide functionality that aids the IR process, the continued need for search systems, and research into their improvement, remains essential.

**INTRODUCTION**

Information retrieval (IR, also known as search) systems are widely used tools for information seeking in biomedicine and health and just about all other aspects of our lives. Search systems such as Google or Bing for general Web searching and PubMed for the biomedical literature put the world's archival knowledge at our fingertips for general and biomedical topics (even if paywalls do not always allow immediate access).

IR systems had been relatively mature applications until late 2022, when any staidness of search systems was upended by the emergence of generally-available generative artificial intelligence (AI) chatbots, based on large language models (LLMs), initially with ChatGPT and soon others to follow. Shortly thereafter came generative AI capabilities added to the two major Web search engines, Microsoft Bing and Google. All of a sudden, searching the Web was transformed in ways that many did not see coming.

I contemplate a good deal about search systems not only because they are the focus of my research(1) but also because I am an academic who does a great deal of teaching and writing. One of my main teaching activities is a widely-subscribed introductory course in biomedical and health informatics, in which I aim to impart not only the big picture of the field, but also to provide a broad array of references to document the knowledge that makes up that big picture. I teach another course on evidence-based medicine, which leads me to think not only about how we find the best evidence to support biomedical and health decisions, but also how we use search systems to find and synthesize that evidence.

My work activities may make me different from the average user of search systems, but they do give me a broad perspective on their use and now about the role(s) that generative AI may play. Decades ago, information scientists elucidated the different types of information needs that users bring to search systems. We may search over different types of content now, and have to deal with challenges that were not thought about in those earlier times, such as misinformation and filtering out true content versus advertising or misinformation. But we still search for information to help answer direct questions or make decisions, or to use information for more integrative tasks, such as writing and teaching.

**INFORMATION NEEDS INFORM USE OF LLMS**

The types of information needs that users bring to IR systems have been studied for decades. Lancaster and Warner defined *subject needs*,(2) which fall into three categories:
- Help in solving a certain problem or making a decision
- Background information on a topic
- Keep up with information in a given subject area

They called the first two subject needs *retrospective information needs*, in that documents already published are sought, while the latter need is called a *current awareness need*, which is met by filtering new documents to identify those on a certain topic. Retrospective needs may also be classified by the amount of information needed:
- A single fact
- One or more documents but less than the entire literature on the topic
- A comprehensive search of the literature

Wilkinson and Fuller described four types of information needs for document collections:(3)
- Fact-finding – locating a specific item of information
- Learning – developing an understanding of a topic
- Gathering – finding material relevant to a new problem not explicitly stated
- Exploring – browsing material with a partially specified information need that can be modified as the content is viewed

Another type of information need is *known-item searching*, where we know what specific information item we are seeking, and now want to find it. Many academics do quite a bit of this latter task, for example wanting to retrieval a scientific paper or preprint when we only have its title or less-than-complete metadata about it.

All of these varied information needs are at odds with the output of generative AI chatbots that provide no or few references. Even when references are provided, they often do not provide a direct citation for what is said. Furthermore, the user has no idea what actual source text led the generative model to output the specific text.

An additional aspect to searching that is critical to academics and many others when seeking information is the authoritativeness of information items retrieved. This is certainly important in scientific areas, especially in biomedicine and health. When searching on biomedical and health-related topics, one of the first things I look at when retrieving something is who wrote it and who published it. This process is not just limited to my academic searching. Even when I am searching on other topics, e.g., current events or consumer products, I want to know the author(s) and source(s) of the information retrieved. I also often find myself trying to trace statements and assertions back to their source. Even for consumer-oriented information, but certainly for academic searching, I want to know what research supports, for example, a given claim that some diagnostic test should be used or some treatment prescribed. Likewise, if someone makes a claim about a research method or results, I often want to find the original research to make sure the claim is supported by someone's interpretation of it.

A final aspect of search that is important to many is timeliness, in that we want to retrieve the most up-to-date information. This is especially critical in rapidly evolving fields, for example, the use of generative AI in biomedical and health applications. The large resource requirements of training LLM models only allows them to be updated intermittently, and they may not reflect the latest information about a topic.

**CHALLENGES FOR LLMS IN SEARCH**

In the early days of the World Wide Web, there was much concern about the "quality" of information on the Web and returned by Web search engines.(4) As the Web democratized publishing, i.e., anyone with access to a Web server could post information, IR researchers thought about ways to aid searchers in identifying the veracity of what they retrieved. A colleague and I, with countless others, thought about methods to automate detection of factors related to the quality on health information.(5) This problem was never really solved, although the emergence of Google helped in one important regard by ranking search output by links to given pages, which was a good if imperfect proxy for quality. Nonetheless, the information quality war has probably been lost, especially with emergence of social media as well as methods for manipulating the retrieval of disinformation.(6,7)

It is from this perspective that we can contextualize the emergence of generative AI, not only as standalone chatbots such as ChatGPT, but also embedded in search systems such as Bing and Google. Clearly there are times when a simple answer, coming out of a search or a chatbot, is sufficient. But it is when my information need goes beyond a simple question that I find LLMs still to be wanting. Whether it is searching for information to make a decision about my personal health or to synthesize an area of science for my teaching or research, I need more than the general commentary of an LLM or its provision of a short list of references or Web sites.

A number of other IR researchers have thought about these issues. Shah and Bender discussed "situating search," noting the varied information-seeking tasks that lead us to use search, and now generative AI, systems.(8) These systems must account for various aspects of searching, including diverse information-seeking needs and strategies, types of searchers, and bias in search results. Shah has also noted some of the challenges for LLM systems in the context of IR systems:(9)
- Opacity and hallucinations – LLMs "don't know when they don't know"
- Stealing content and Web site traffic – LLMs "learn from other people's content and may divert traffic from their Web sites"
- Taking away learning and serendipity – "search is exploring and we may learn new unrelated things"

One additional concern in modern times about chatbots replacing search is the increased energy consumption of the former, especially in an era of concern over climate change. AI systems not only use a great deal of electricity to power servers for training large models, but they also use more energy during user interactions. One recent study estimated a Google search using its generative AI capabilities consumed ten times more energy than a plain Google search.(10)

**FUTURE ROLE OF LLMS IN SEARCH**

There has been some early research to look at information seeking and search system use in the LLM era, but there is not yet any sort of comprehensive picture of the role of chatbots vs.

search for different information needs. A few studies have noted that ChatGPT does not provide comprehensive references and may even confabulate them in areas such as learning health systems,(11) cryptococcal meningitis,(12) and a variety of other interdisciplinary topics.(13)

It may be possible that LLM systems can augment the search process, but the evidence so far is slim. One study looked at the potential for generating Boolean queries for systematic review search and found improved precision but at a cost to recall, which is critically important in such searches.(14) Another study found that ChatGPT could be help in answering consumer-health questions but correctness reduced when prompting included supporting evidence in the form of documents).(15) There may also be potential value for methods such as retrieval-augmented generation, where search engine output is used to prompt existing LLMs for more focused and up-to-date information.(16) Additional value may emanate from improved methods of prompting(17) or the addition of knowledge graphs to augment the prompting process.(18) At this time, however, there is little experimental evidence for the value of these methods.

These are still early days for LLMs and generative AI in search, and there may yet be development of the critical features we desire for IR systems that are noted above. ChatGPT and the generative AI in Bing and Google are fascinating to explore, and require the attention of everyone who may employ future iterations of them in health-related settings. But for critical information needs of academics like myself, whether seeking information about personal aspects of life or scholarly work, I still scroll past the generative AI at the top of Bing and Google to the search functions while heading to sites like PubMed to seek scientific literature. Whether I retrieve a news article, a commentary, or a scientific paper, I still look for who wrote the article and in which venue it was published. I then delve into the text looking for the information I need to decide whether this item is useful for the task(s) that led me to retrieve it.

As such, I still head first to search engines over generative AI chatbots for information seeking. As I prepare lectures, papers, and other intellectual syntheses, who wrote the paper, report, news story, etc. and where it was published are as important as the content itself. ChatGPT and other chatbots produce interesting information, but I find it less valuable for my work than its original source. This may change in the future as LLM systems become more powerful and trained to searching use cases, although we also know that the history of biomedical and health informatics is littered with applications that had great hype and never achieved the revolutionary use that was anticipated.

## REFERENCES

1. Hersh W. Information Retrieval: A Biomedical and Health Perspective. 4th ed. Springer International Publishing; 2020.

2. Lancaster FW, Warner AJ. Information retrieval today. Rev., retitled, and expanded ed. Arlington, Va.: Information Resources Press; 1993.

3. Wilkinson R, Fuller M. Integration of information retrieval and hypertext via structure. In: Agosti M, Smeaton A, editors. Information Retrieval and Hypertext. Springer; 1996.

4. Silberg WM, Lundberg GD, Musacchio RA. Assessing, controlling, and assuring the quality of medical information on the Internet: Caveant lector et viewor--Let the reader and viewer beware. JAMA. 1997 Apr 16;277(15):1244–5.

5. Price SL, Hersh WR. Filtering Web pages for quality indicators: an empirical approach to finding high quality consumer health information on the World Wide Web. Proc AMIA Symp. 1999;911–5.

6. Boyd D, Golebiewski M. Data & Society. Data & Society Research Institute; 2019 [cited 2022 Apr 22]. Data Voids. Available from: https://datasociety.net/library/data-voids/

7. Center for Countering Digital Hate [Internet]. 2021 [cited 2021 Sep 28]. The Disinformation Dozen. Available from: https://www.counterhate.com/disinformationdozen

8. Shah C, Bender EM. Situating Search. In: Proceedings of the 2022 Conference on Human Information Interaction and Retrieval [Internet]. New York, NY, USA: Association for Computing Machinery; 2022. p. 221–32.

9. Shah C. The Conversation. 2023 [cited 2023 Aug 22]. AI information retrieval: A search engine researcher explains the promise and peril of letting ChatGPT and its cousins search the web for you. Available from: http://theconversation.com/ai-information-retrieval-a-search-engine-researcher-explains-the-promise-and-peril-of-letting-chatgpt-and-its-cousins-search-the-web-for-you-200875

10. de Vries A. The growing energy footprint of artificial intelligence. Joule. 2023 Oct 18;7(10):2191–4.

11. Chen A, Chen DO. Accuracy of Chatbots in Citing Journal Articles. JAMA Netw Open. 2023 Aug 1;6(8):e2327647.

12. Schwartz IS, Link KE, Daneshjou R, Cortés-Penfield N. Black Box Warning: Large Language Models and the Future of Infectious Diseases Consultation. Clin Infect Dis Off Publ Infect Dis Soc Am. 2023 Nov 16;ciad633.

13. Walters WH, Wilder EI. Fabrication and errors in the bibliographic citations generated by ChatGPT. Sci Rep. 2023 Sep 7;13(1):14045.

14. Wang S, Scells H, Koopman B, Zuccon G. Can ChatGPT Write a Good Boolean Query for Systematic Review Literature Search? arXiv; 2023. Available from: http://arxiv.org/abs/2302.03495

15. Zuccon G, Koopman B. Dr ChatGPT, tell me what I want to hear: How prompt knowledge impacts health answer correctness. arXiv; 2023. Available from: http://arxiv.org/abs/2302.13793

16. Lewis P, Perez E, Piktus A, Petroni F, Karpukhin V, Goyal N, et al. Retrieval-augmented generation for knowledge-intensive NLP tasks. In: Proceedings of the 34th International Conference on Neural Information Processing Systems. Red Hook, NY, USA: Curran Associates Inc.; 2020. p. 9459–74.

17. Nori H, Lee YT, Zhang S, Carignan D, Edgar R, Fusi N, et al. Can Generalist Foundation Models Outcompete Special-Purpose Tuning? Case Study in Medicine. arXiv; 2023. Available from: http://arxiv.org/abs/2311.16452

18. Pan S, Luo L, Wang Y, Chen C, Wang J, Wu X. Unifying Large Language Models and Knowledge Graphs: A Roadmap. arXiv; 2023. Available from: http://arxiv.org/abs/2306.08302